\documentclass[a4paper,fleqn,usenatbib]{mnras}

\usepackage{newtxtext,newtxmath}

\usepackage[T1]{fontenc}
\usepackage{ae,aecompl}

\usepackage{amsmath}	
\usepackage{booktabs}
\usepackage[font=small,labelfont=bf]{caption}
\usepackage{color}
\usepackage{graphicx}	
\usepackage{mathtools}
\usepackage{subfigure}
\usepackage{systeme}
\usepackage{ulem}

\DeclareMathOperator{\sech}{sech}
\newcommand{\DE}[1]{#1}

\definecolor{purple}{rgb}{0.5, 0.0, 0.5}

\renewcommand{\vec}[1]{\mathbf{#1}}

\title[...]{The impact of resistive electric fields on particle acceleration in reconnection layers}

\author[E. Puzzoni et al.]{
E. Puzzoni$^{1}$\thanks{E-mail: eleonora.puzzoni@edu.unito.it}, A. Mignone$^{1}$ and G. Bodo$^{2}$
\\
$^{1}$Physics Department, Turin University, Via Pietro Giuria 1, 10125 Torino, Italy\\
$^{2}$INAF – Osservatorio Astrofisico di Torino, Strada Osservatorio 20, I-10025 Pino Torinese, Italy
}

\date{Accepted XXX. Received YYY; in original form ZZZ}

\pubyear{2015}

\begin{document}
\label{firstpage}
\pagerange{\pageref{firstpage}--\pageref{lastpage}}
\maketitle

\begin{abstract}

\noindent In the context of particle acceleration in high-energy astrophysical environments featuring magnetic reconnection, the importance of the resistive term of the electric field compared to the convective one is still under debate. 
In this work, we present a quantitative analysis through 2D magnetohydrodynamic numerical simulations of tearing-unstable current sheets coupled to a test-particles approach, performed with the PLUTO code.
We find that the resistive field plays a significant role in the early-stage energization of high-energy particles.
Indeed, these particles are firstly accelerated due to the resistive electric field when they cross an X-point, created during the fragmentation of the current sheet.
If this preliminary particle acceleration mechanism dominated by the resistive field is neglected, particles cannot reach the same high energies.
Our results support therefore the conclusion that the resistive field is not only non-negligible but it does actually play an important role in the particle acceleration mechanism.


\vspace{4mm}

\vspace{4mm}
\end{abstract}

\section{Introduction}
\label{sec:intro}
The study of particle acceleration in astrophysical plasmas such as gamma-ray bursts \citep[GRBs; see, e.g.,][]{Giannios2008, Zhang2011, Sironi2013, Sironi&Giannios2013, Beniamini2014, Beniamini2017}, blazar jets \citep[see, e.g.,][]{Giannios2013, Sironi2015, Petropoulou2016}, supernova remnants \citep[SNRs; see, e.g.,][]{Bell2013, Morlino2013, Caprioli2014} and pulsar wind nebulae \citep[PWNe; see, e.g.,][]{Bucciantini2011,Sironi2011,Cerutti2012,Cerutti2014} is important for the interpretation of the observed spectra of these astrophysical objects. 
Indeed, the investigation of the particle acceleration process would allow a much better understanding of the high energy emission from these objects.

These studies are carried out through numerical simulations, for which several models are available.
On the one hand, Particle-In-Cell (PIC) methods provide the most comprehensive and consistent plasma description at kinetic scales but are inherently limited by resolution constraints fixed by the ion or electron inertial lengths
\citep[see, e.g.,][]{Zenitani2001, Zenitani2008, Jaroschek2004, Lyubarsky2008, Oka2010, Sironi2014, Guo2015, Cerutti2014}.
On the other hand, the magnetohydrodynamic (MHD) model is appropriate for larger scales (i.e., well beyond the particle gyroradius or inertial length) and can be easily coupled to a test-particles approach to investigate acceleration mechanisms \citep[see, e.g.,][]{Onofri2006, Kowal2011, Kowal2012, Zhou2016, Ripperda2017a, Ripperda2017b, Puzzoni2021}.
Test-particles are affected by the electromagnetic field of the plasma but, in turn, do not exert any force on it (no back-reaction), and are often evolved on a fluid snapshot \citep[i.e., where the fluid is frozen in time see, e.g.,][]{Onofri2006, Kowal2011, Kowal2012, Ripperda2017a}.
In this work, conversely, we evolve test-particles along with the fluid \citep[as in, e.g.,][]{Ripperda2017b, Puzzoni2021}.

Of the various acceleration mechanisms, magnetic reconnection is thought to be the most efficient at high magnetizations as it produces power-law spectra of energetic non-thermal particles  \citep[see, e.g.,][]{Zenitani2001, Jaroschek2004, Onofri2006, Bessho2007, Lyubarsky2008, Guo2014, Li2015, Werner2016}.
During the magnetic reconnection process, magnetic energy is converted into thermal and kinetic energy of the plasma as the magnetic field lines of opposite polarity annihilate. 
Magnetic reconnection leads to efficient acceleration when it is triggered by dynamical instabilities such as the tearing mode, during which an initially neutral current sheet fragments into X- and O-points, or plasmoids \citep{Loureiro2016}. 
In the X-points, the magnetic field has null points, while the O-points are characterized by a high current density. 

The orbits followed by the accelerated particles inside the reconnection layer are mainly of the Speiser type \citep[see][]{Speiser1965}, which sample both sides of the current sheet \citep[see, e.g.,][]{Zenitani2001, Giannios2010, Cerutti2013, Zhang2021}.
Particles are accelerated by the plasma electric field, which consists of a convective and a resistive term associated with different acceleration mechanisms. 
Particles can be in fact accelerated directly by the resistive electric field at X-points \citep[see, e.g.,][]{Zenitani2001, Bessho2007, Lyubarsky2008, Sironi2014, Nalewajko2015, Ball2019} or
at the secondary current sheet that forms at the interface between two merging plasmoids \citep[see, e.g.,][]{Oka2010,Sironi2014,Nalewajko2015}.
The energization process can also occur when particles are trapped in a contracting plasmoid due to the Fermi reflection \citep[see, e.g.,][]{Drake2006,Drake2010,Kowal2011,Bessho2012, Petropoulou2018,Hakobyan2021}, associated to the convective electric field.

Which of the two electric field terms dominates in accelerating particles is to date still under debate.
\cite{Kowal2011} and other authors \citep[see, e.g.,][]{Kowal2012,deGouveia2015,delValle2016,Medina2021}, for example, directly neglect the contribution of the resistive field in the particle acceleration mechanism, as they consider it unimportant.
In \cite{Guo2019} and \cite{Paul2021} it is claimed that the Fermi mechanism is the dominant one in the acceleration process, while the crossing of an X-point makes a small contribution to the global energization. 
In particular, \cite{Guo2019} argue that the non-ideal field does not contribute even to the formation of the power-law, but it is only the Fermi mechanism that determines the spectral index.

On the contrary, \cite{Onofri2006} and \cite{Zhou2016} argue that, even if the resistive contribution is less intense than the convective one, it is much more important in accelerating particles. 
The results of \cite{Zhou2016} were later confirmed by \cite{Ripperda2017a}.
In addition, in \cite{Ball2019} it is claimed that the particle acceleration mechanism is more efficient when more X-points are formed.
Recently, \cite{Sironi2022} claimed that the acceleration from the non-ideal electric field is a basic requirement for subsequent acceleration, which is on the contrary typically dominated by the ideal field \citep[as argued by][]{Guo2019}. 
In fact, the particles that do not undergo the non-ideal contribution do not even reach relativistic energies. 

While in \cite{Puzzoni2021} we focused on the problem of convergence concerning the numerical method, here we aim at quantifying the relative importance of the resistive and convective electric field in the particle acceleration process to understand if, when, and why one contribution prevails over the other.
The paper is organized as follows. 
In Section \ref{sec:model} the theoretical and numerical model is explained.
The results are discussed in Section \ref{sec:particle_acceleration}, divided in preliminary considerations made with 2D histograms (\ref{sec:2D_hist}), the study on the importance of the resistive field in accelerating high-energy particles (\ref{sec:res_role}), and the one on the impact of the current sheet evolutionary phases on particle energization (\ref{sec:cs}).
A summary is given in Section \ref{sec:summary}.

\section{Theoretical and numerical model}
\label{sec:model}
Our model follows essentially the same configuration adopted by \cite{Puzzoni2021}, consisting of a perturbed Harris sheet, whose profile is defined by
\begin{equation}
  B_x (y) = B_0 \tanh \left(\frac{y}{a}\right),
  \label{eq:B}
\end{equation}
where $a = 250 \  c / \omega_p$ denotes the initial width of the current sheet (here $\omega_p$ indicate the plasma frequency).
We normalize the magnetic field strength $B_0$ such that our unit velocity is the Alfv{\'e}n velocity $v_A$, so $\rho_0 = B_0 = 1$ in code units. 
The guide field is absent ($B_z = 0$), and the initial equilibrium is achieved by balancing the Lorenz-force term with a thermal pressure gradient, where the plasma $\beta$ parameter is set to $0.01$ outside the current sheet region,  
\begin{equation}
    p(y) = \frac{1}{2} B_0^2 (\beta + 1) - \frac{1}{2} B_x^2 (y)\,.
\end{equation}

Resistive instabilities are triggered by introducing a fixed number of small-amplitudes modes with different wavenumbers $k$.
We achieve this more conveniently by redefining the vector potential according to  $A_z(x,y) = A_0(y) + \delta A_z(x,y)$ where $A_0(y)= a B_0 \log(\cosh(y/a))$ corresponds to the equilibrium magnetic field (equation \ref{eq:B}), while the perturbed term is defined as
\begin{equation}
    \delta A_{z}(x,y) = \frac{\epsilon B_0}{N_m} \sum_{m = 0}^{N_m} \frac{1}{k_m} \sin(k_mx + \phi_m) \sech\left(\frac{y}{a}\right),
\end{equation}
where $\epsilon = 10^{-3}$ is the perturbation amplitude, $N_m$ is the number of modes \citep[20, as in][]{Puzzoni2021}, $k_m=(m+1)k_0 = 2 \pi (m+1)/L$ are the wavenumbers, and $\phi_m$ are random phases.

We solve the equations of resistive non-relativistic MHD as described in \cite{Puzzoni2021} using the PLUTO code \citep[see][]{Mignone2007, Mignone2012} with the 5$^{\mathrm{th}}$-order WENO-Z reconstruction algorithm \citep[see, e.g.,][]{Borges2008,Mignone2010} in combination with the HLLD Riemann solver of \cite{Miyoshi2005} and the UCT-HLLD emf averaging scheme. 
The redefined Lundquist number $\bar{S}=v_A a/\eta$, corresponding to the current sheet's width $a$, is set to $10^4$. 
The 2D domain is rectangular of size $L \times L/2$ (where $L = 2 \times 10^4 \ c/\omega_p$). 
In the MHD equations, the actual speed of light does not explicitly appear, therefore the artificial value $\mathbb{C} = 10^4 \ v_A$ is used as we want a fluid velocity not comparable to the speed of light. The chosen value of $\mathbb{C}$ can be representative of some astrophysical environments such as coronal mass ejections \citep[CMEs; see, e.g.,][]{Maguire2020} and solar wind \citep[SW; see, e.g.,][]{Bourouaine2012}.
The grid resolution is set to $1536 \times 768$ (i.e. $a/\Delta x \sim 20$) and then progressively doubled up to $6144 \times 3072$ (i.e. $a/\Delta x \sim 80$), in order to determine convergence even in the non-linear evolution phase of the current sheet. 

Test-particles are initialized on the grid one per cell and their initial velocities follow a Maxwellian distribution.
Particles obey the equations of motion
\begin{equation}\label{eq:particles}
 \begin{dcases} 
  \frac{d \textbf{x}_p}{d t} = \textbf{v}_p \\ 
  \frac{d (\gamma \textbf{v})_p}{d t} = \left(\frac{e}{mc}\right)_p (c \textbf{E} + \textbf{v}_p \times \textbf{B}) \,,
  \end{dcases}
\end{equation}
where $\textbf{x}_p$ and $\textbf{v}_p$ represent the spatial coordinate and velocity respectively, $\gamma= (1 - \textbf{v}^2_p/\mathbb{C}^2)^{-1/2}$ is the Lorentz factor while $(e/mc)_p$ is the particle charge to mass ratio. 
The suffix $p$ will be used to label a generic particle.
For computational reasons, the mass of the particles composing the fluid is set equal to that of the accelerated particles.
Consequently, the charge to mass ratio in equation (\ref{eq:particles}) becomes unity when written in code units.

Equation (\ref{eq:particles}) is solved using a standard Boris pusher \citep[see][]{Mignone2018} and the magnetic and electric fields, $\vec{B}$ and $\vec{E}$, are obtained from the fluid.
More specifically, the electric field in resistive MHD is defined by
\begin{equation} \label{eq:electricfield}
  c \textbf{E} = - \textbf{v}_g \times \textbf{B} + \frac{\eta}{c} \textbf{J},
\end{equation}
where the first term is the convective term ($\textbf{E}_c$) while the second one corresponds to the resistive electric field ($\textbf{E}_r$), where $\eta$ represents the scalar resistivity, assumed constant in this work.
The current density is defined as usual by
\begin{equation}\label{eq:currentdensity}
  \textbf{J} = c\nabla \times \textbf{B}
\end{equation}
(notice that a constant factor of $\sqrt{4\pi}$ has been reabsorbed in the definition of $\vec{B}$).

In this work, we focus precisely on the relative importance of these two terms in the particle acceleration mechanism.
Each contribution exerts a work on the particles with a corresponding change in their kinetic energy $E_\mathrm{kin} = (\gamma-1)\mathbb{C}^2$.
During a single time step $\Delta t^n$ this variation is given by the sum of the two contributions
\begin{equation}
    \Delta E^n_\mathrm{kin} = h^nc
    \left(\textbf{E}_c \cdot \textbf{v}_p + \textbf{E}_r \cdot \textbf{v}_p\right)^n,
    \label{eq:work}
\end{equation}
where $h^n =  \Delta t^n(e/mc)_p$ is related to the simulation time step $\Delta t^n$.
In what follows we will make use of the energy gained by the particles which, following equation (\ref{eq:work}), will split into
\begin{equation}
  W =   \sum_n \left(h c \vec{E}_c \cdot \vec{v}_p \right)^n
      + \sum_n \left(h c \vec{E}_r \cdot \vec{v}_p\right)^n \equiv W_c + W_r,
\end{equation}
where $n$ indicates the step number.
From now on, (specific) energy and work $W$ will be normalized to $v_A^2$, where $v_A=B_0/\sqrt{\rho_0}$ is the nominal initial Alfv{\'e}n velocity.

\section{Results from Numerical Simulations}
\label{sec:particle_acceleration}

\begin{figure*}
    \centering
    \includegraphics[width=0.33\textwidth]{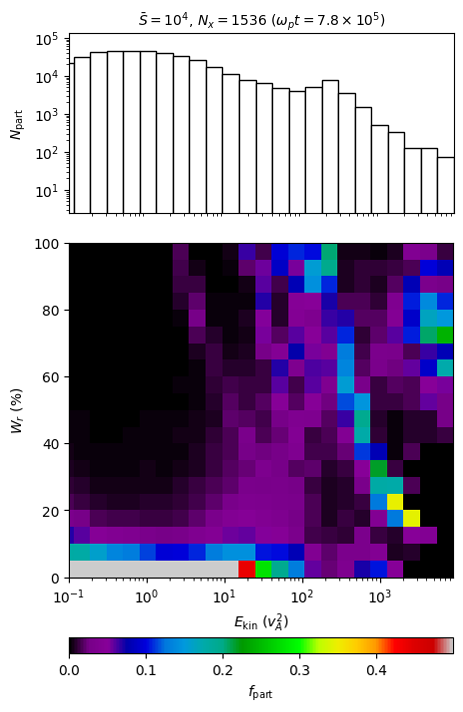}%
    \includegraphics[width=0.33\textwidth]{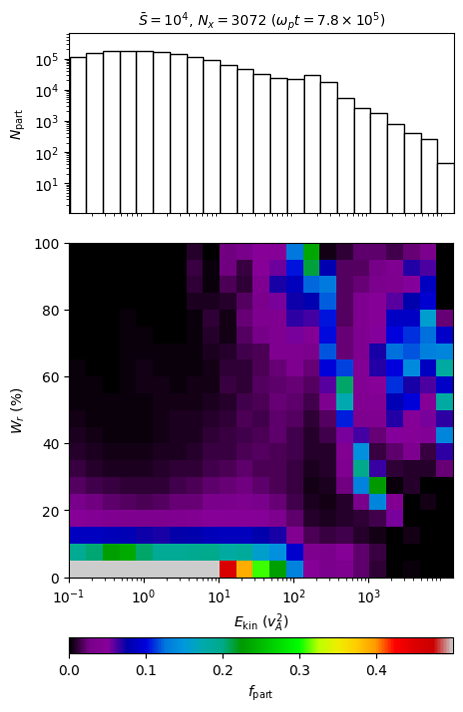}%
    \includegraphics[width=0.33\textwidth]{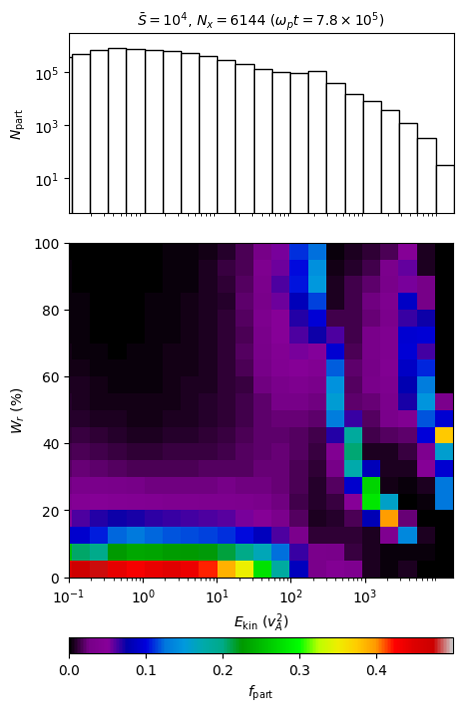}  
    \caption{ 2D histograms of the energy gained (in $\%$) due to the resistive electric field by particles as functions of their kinetic energy at the end of the computational time ($\omega_p t = 7.8 \times 10^5$). The colorbar represents the fraction of particles in each energy bin. The total number of particles in each bin is shown in the corresponding upper panels. These histograms are reproduced at the $N_x = 1536$ (left panels), $N_x = 3072$ (middle panels), and $N_x = 6144$ (right panels) grid resolutions, with $\bar{S} = 10^4$.}
    \label{fig:Hist_2D}
\end{figure*}

We performed simulations with $\bar{S}=10^4$ at different grid resolutions to check the convergence properties. 
Moreover, to investigate the importance of the resistive field, numerical computations are carried out twice by first including and then removing the resistive contribution from the equation of motion of the particles, while always keeping it during the fluid evolution.
This means that the fluid evolution is the same in both cases while only the particle evolution differs.


\subsection{Preliminary Considerations: 2D Histogram Analysis}
\label{sec:2D_hist}
While in \cite{Puzzoni2021} we investigated convergence by concentrating on the particle spectra, in this paper, we shift our focus on the convergence properties of the resistive field, by investigating its behaviour at different grid resolutions.
Indeed, a detailed analysis of the resistive contribution is provided by the histograms in Figure \ref{fig:Hist_2D} at the resolution of $N_x = 1536, 3072$ and $6144$ grid zones.
The 2D histograms show the percentage of energy gained by particles due to the resistive electric field at the end of the computational time ($\omega_p t = 7.8 \times 10^5$), as a function of their final kinetic energy.
Notice that a steady state is reached at this time \citep[called \lq\lq saturation phase\rq\rq\ in][]{Puzzoni2021}.
The colors indicate the fraction of particles ($f_\mathrm{part}$) in each energy bin, normalized to the total number of particles in that bin ($N_\mathrm{part}$), which in turn is reported in the corresponding uppermost panels.

As an illustrative example, it is worth looking at the first gray pixel in the first energy bin (lower left corner of the figure) of the middle 2D histogram of Figure \ref{fig:Hist_2D}.
$50\%$ of particles in this energy bin ($N_\mathrm{part} \approx 10^5$, from the corresponding upper panel) is energized between $0-5\%$ by the resistive field.
The corresponding 2D histogram for the convective contribution would be mirrored with respect to the $x$-axis.
Therefore, $50\%$ of particles in the example bin is energized between $95-100\%$ by the convective electric field.

By looking at Fig. \ref{fig:Hist_2D}, the 2D histograms at different grid resolutions show a similar shape. 
The resistive electric field has a small contribution ($W_r \lesssim 30\%$) at low energies ($E_\mathrm{kin} \lesssim 10$).
On the contrary, it has a non-negligible contribution at intermediate energies ($10 \lesssim E_\mathrm{kin} \lesssim 10^3$).
Indeed, up to about $30\%$ of particles are energized up to $100\%$ by the resistive field (see the purple, blue and green pixels).
At high-energies ($E_\mathrm{kin} \gtrsim 10^3$), the resistive contribution slightly decreases with the grid resolution, but still leaves the corresponding energy gain significant.
Indeed, even in the high-resolution case (see rightmost panel), for about $40\%$ of the particles in the last energy bin, the resistive contribution accounts for $40-45\%$ of their energy gain (see orange pixel).
A smaller percentage of particles ($\lesssim 10\%$) are accelerated up to $100\%$ by the resistive field (see purple pixels in the upper right corner).
In the low-resolution case (see leftmost panel), owing to increased numerical diffusion, the amount of work exerted by the resistive electric field is somewhat larger at high-energies ($W_r \gtrsim 50\%$, purple, blue, and green pixels).

Our simulation results indicate that the contribution of the resistive electric field converges with grid resolution at intermediate energies ($10 \lesssim E_\mathrm{kin} \lesssim 10^3$). 
At high energies ($E_\mathrm{kin} \gtrsim 10^3$), however, convergence assessment is somewhat more uncertain, as the contribution of the resistive field appears to decrease as resolution increases.
Further details on plasma convergence are presented in Appendix \ref{app:plasma_convergence}, where we conclude that it was not possible - with the current computational resources - to resolve the typical critical current sheet \citep[i.e., the smallest current layer found in the system; see][]{Uzdensky2010} in the fragmentation phase.
Nevertheless, the resistive contribution is non-negligible at all resolutions. 
Indeed, in the following sections, we will assess its fundamental importance in the particle acceleration mechanism.

\subsection{The role of the resistive field on high-energy particles}
\label{sec:res_role}
\begin{figure*}
    \centering
    \includegraphics[width=0.33\textwidth]{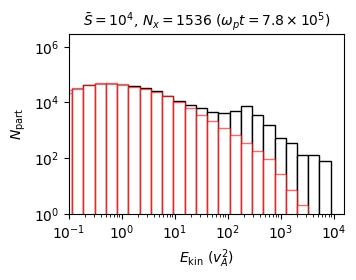} 
    \includegraphics[width=0.333\textwidth]{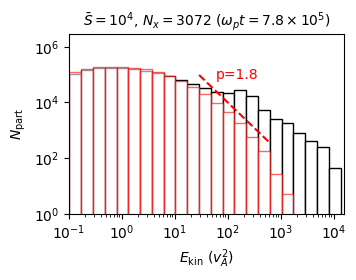} 
    \includegraphics[width=0.33\textwidth]{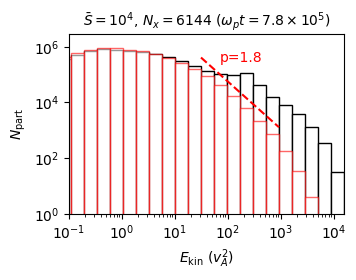}
    \caption{Histograms of the particles kinetic energy at the end of the computational time ($\omega_p t = 7.8 \times 10^5$) for $\bar{S}=10^4$ at the $N_x = 1536$ (left panel), $N_x = 3072$ (middle panel), and $N_x = 6144$ (right panel) grid resolutions, obtained with (black bars) and without (red bars) the resistive term in the particle equation of motion.}
    \label{fig:hist_1D_br}
\end{figure*}

Fig. \ref{fig:hist_1D_br} shows the 1D histograms of the particles kinetic energy at the end of the computational time ($\omega_p t = 7.8 \times 10^5$) for $\bar{S}=10^4$ at different grid resolutions ($N_x = 1536, 3072, 6144$).
The black and the red bars are obtained, respectively, by including or excluding the resistive term in the particle equations of motion.

Following the work of \cite{Sironi2022}, we have decided to remove all the particles that are initially found within the current sheet ($|y_0/a| < 1$) as they have peculiar behavior depending on the initial conditions. 
Moreover, we removed the particles initially farthest from the current sheet ($|y_0/a| > 16$), as they will never reach it within the final simulation time.
This avoids a box size effect or a statistic bias. 
A direct comparison between the two cases clearly indicates that, when the resistive contribution is neglected, the particle spectra are somewhat steeper, characterized by a power-law with index $p \approx 1.8$ at the largest resolutions ($N_x = 3072, 6144$), and the maximum kinetic energy achieved by the particles is lower.
This discrepancy occurs at all resolutions considered here.
For instance, in the $N_x=3072$ case, the maximum kinetic energy achieved by particles by considering the resistive term is $E_\mathrm{kin} \approx 1.3 \times 10^4$, versus the maximum kinetic energy $E_\mathrm{kin} \approx 1.4 \times 10^3$ achieved without including this term.
Similarly, in the $N_x = 6144$ case, the maximum kinetic energy reached by particles is lower when the resistive term is neglected ($E_\mathrm{kin} \approx 4.2 \times 10^3$ versus $E_\mathrm{kin} \approx 1.6 \times 10^4 $).
This confirms, as also argued by \cite{Sironi2022}, that the resistive electric field contribution is fundamental in building the high-energy tail.
As we shall see shortly, this effect takes place in the early stages of the acceleration process.

\begin{figure*}
    \centering
    \includegraphics[width=0.49\textwidth]{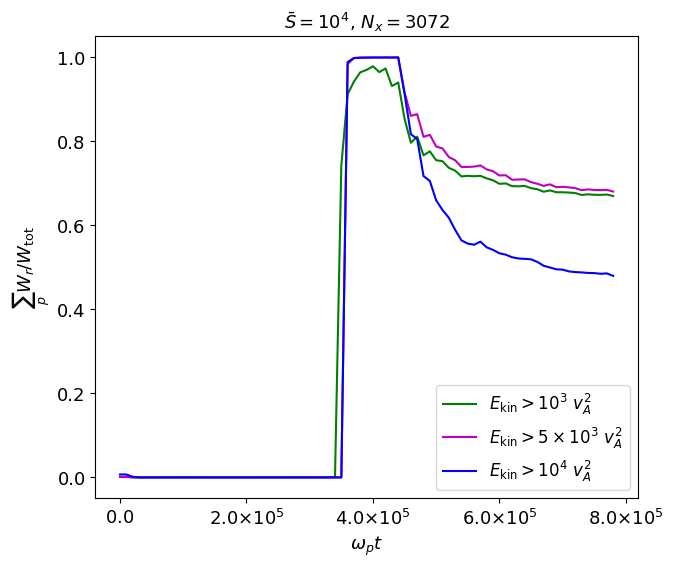}
    \includegraphics[width=0.49\textwidth]{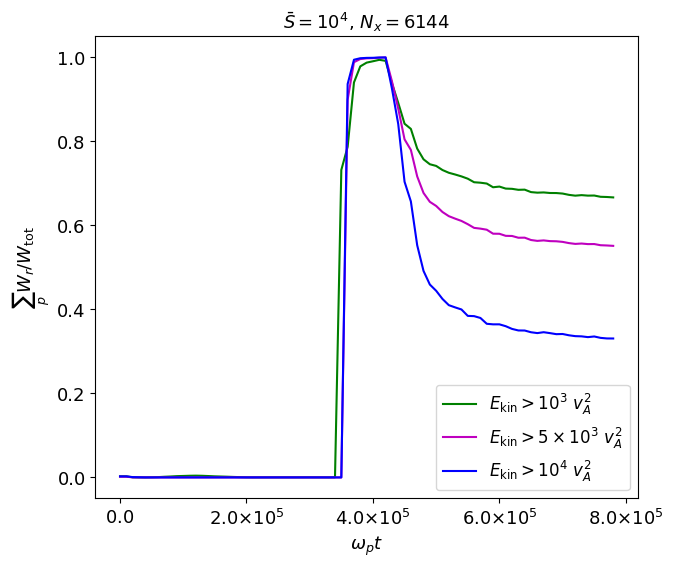}
    \caption{ \textit{Left panel}: Resistive contribution over time on particles that at the end of the computational time reach a kinetic energy of $10^3 \ v_A^2$ (green line), $5 \times 10^3 \ v_A^2$ (magenta line), and $10^4 \ v_A^2$ (blue line) for the $\bar{S}= 10^4$ case with a grid resolution $N_x = 3072$. \textit{Right panel}: Same but for $N_x = 6144$. }
    \label{fig:Wr_tot_conf}
\end{figure*}

To this end, we now focus on the resistive contribution in accelerating particles over time and consider only those particles that at the end of the computational time achieved a kinetic energy $E_\mathrm{kin} > E_\mathrm{thr}$. 
We set three different energy threshold: $E_\mathrm{thr} = 10^3, 5 \times 10^3, 10^4$.
For these particles (indicated by the suffix $p$) we calculated the fraction of energy gained due to the resistive contribution, namely $\sum_p W_\mathrm{r, p}/W_\mathrm{tot}$, where $W_\mathrm{tot} = \sum_p W_\mathrm{r,p} + \sum_p W_\mathrm{c,p}$ is the total energy gained by these particles (i.e., also due to the convective contribution).
Fig. \ref{fig:Wr_tot_conf} shows the fractional energy gain of selected particles due to the action of the resistive field as a function of time, for different energy thresholds and for $\bar{S}=10^4$ with a grid resolution $N_x = 3072$ (left panel) and $N_x = 6144$ (right panel).
By looking at this figure, it is clear that the resistive contribution is dominant in accelerating particles at $3.6 \times 10^5 \lesssim \omega_p t \lesssim 4.4 \times 10^5$ for all the kinetic energy thresholds.
Subsequently, the resistive contribution gradually decreases.
In the $N_x = 6144$ case, the resistive contribution towards the end of computational time is lower (see $E_\mathrm{kin} > 5 \times 10^3 \ v_A^2$ and $E_\mathrm{kin} > 10^4 \ v_A^2$).
This result is in agreement with the decline at high-energies observed in the rightmost panel of Figure \ref{fig:Hist_2D}.
Accordingly, if the resistive contribution is removed from the particle equations of motion, particles cannot achieve the same high energies.

\subsection{Relation between particle energization and current sheet evolution}
\label{sec:cs}
\begin{figure*}
    \centering
    \includegraphics[width=0.45\textwidth]{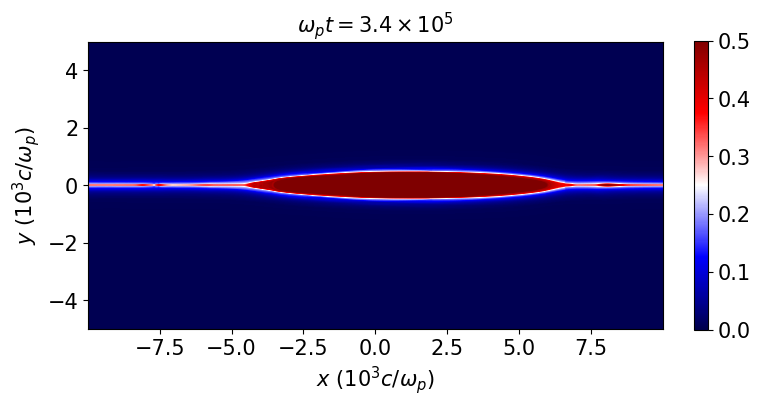} 
    \includegraphics[width=0.45\textwidth]{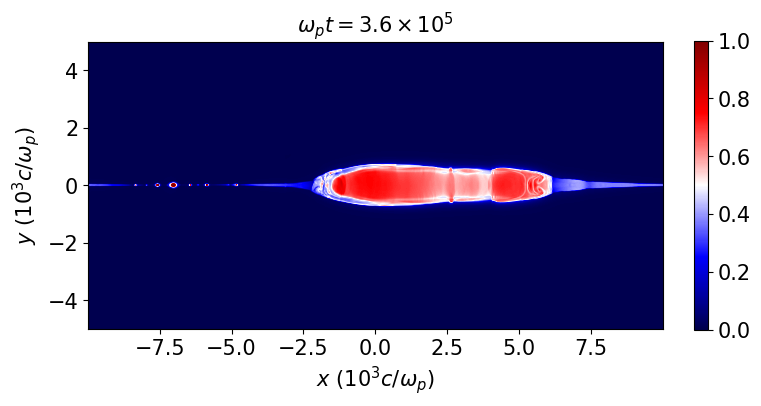} \\
    \includegraphics[width=0.45\textwidth]{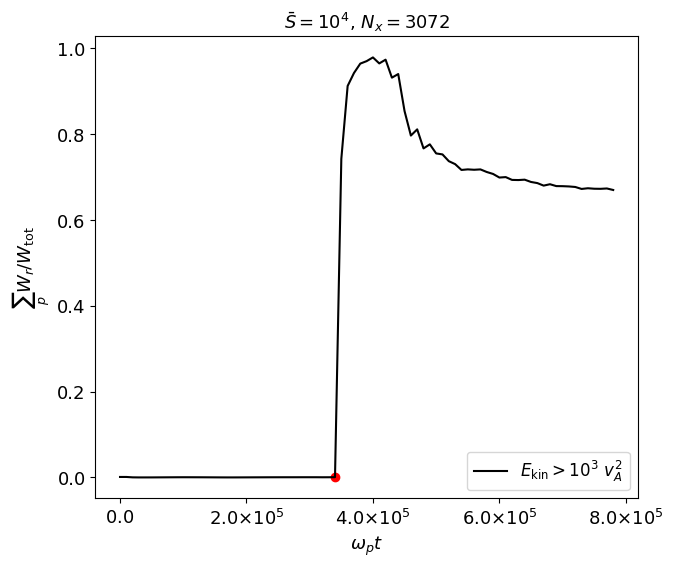} 
    \includegraphics[width=0.45\textwidth]{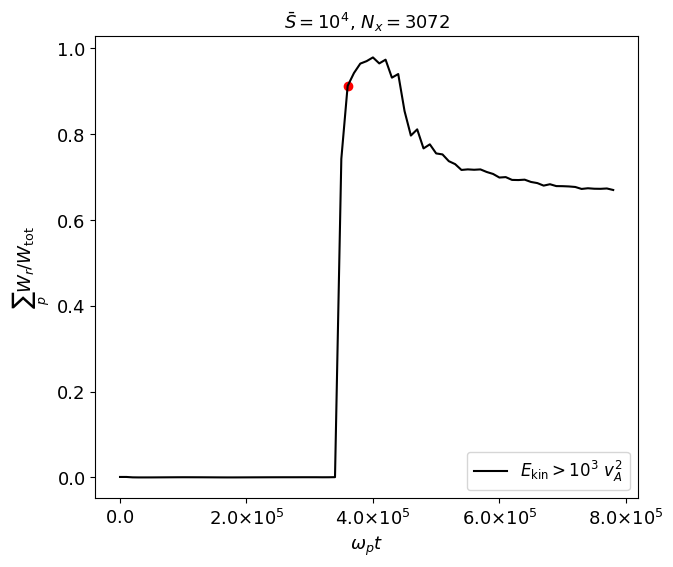} \\
    \includegraphics[width=0.45\textwidth]{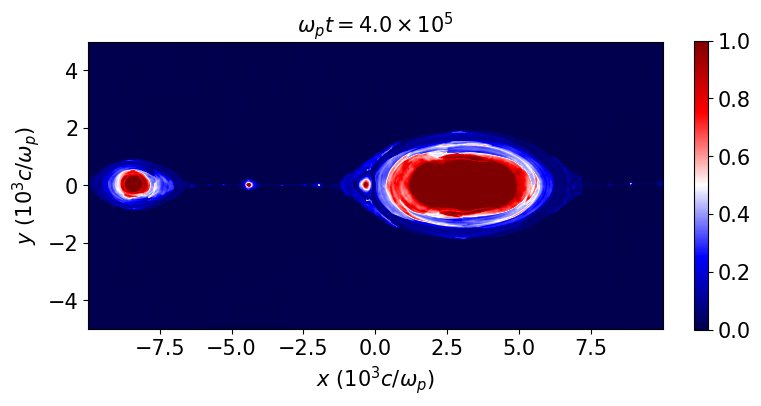} 
    \includegraphics[width=0.45\textwidth]{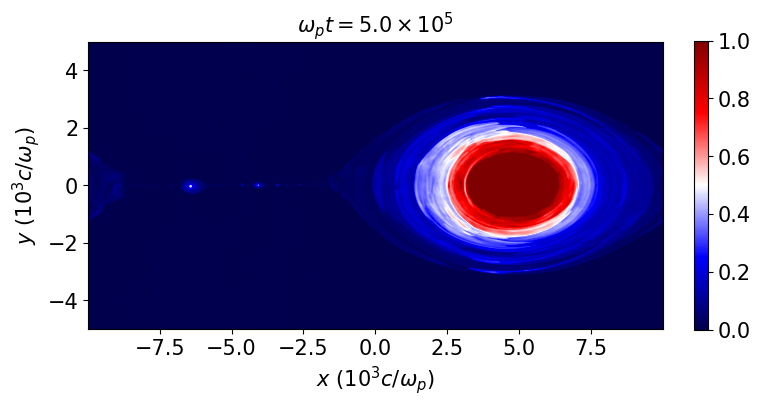} \\
    \includegraphics[width=0.45\textwidth]{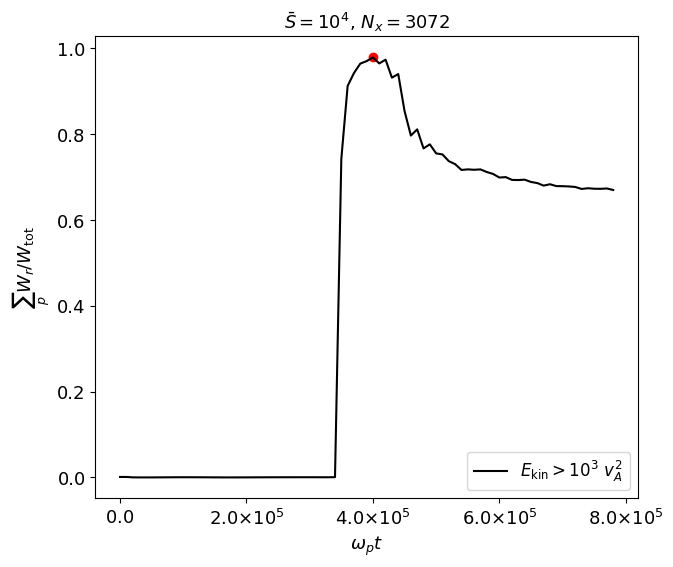}
    \includegraphics[width=0.45\textwidth]{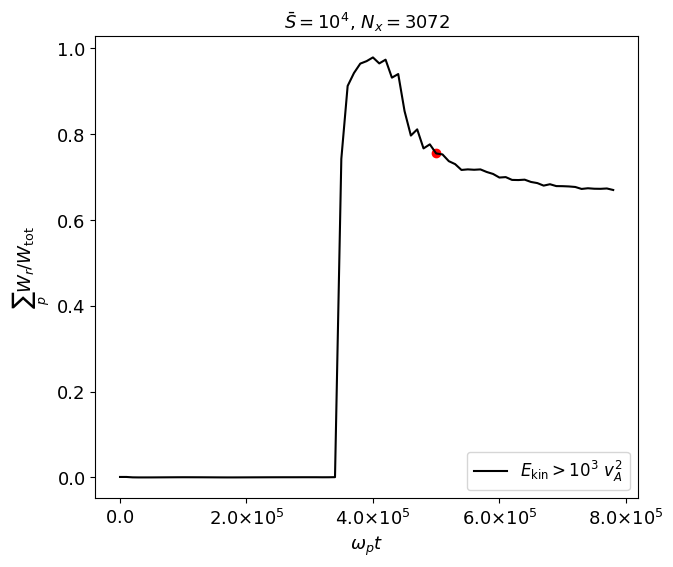}
    \caption{Resistive field contribution on the most energetic particles ($E_\mathrm{kin} > 10^3$) as a function of time, obtained with $\bar{S}=10^4$ and a grid resolution $N_x = 3072$. This plot is repeated four times being marked with red points, that characterized four evolutionary phases of the current sheet, whose pressure is shown in the corresponding upper panels.}
    \label{fig:current_sheet}
\end{figure*}

Fig. \ref{fig:current_sheet} shows the fluid pressure colored maps at four specific times.
These instants are marked with red points in corresponding lower panels, where we show the resistive field contribution for the particles with final kinetic energies above $E_{\rm thr} =10^3$. 
By looking at the upper left panels ($\omega_p t = 3.4 \times 10^5$), it is clear that the current sheet has reached - in the notations of \cite{Puzzoni2021} - the $2^\mathrm{nd}$- linear phase.
At subsequent times ($\omega_p t = 3.6 \times 10^5$, for instance), the current sheet fragments in X- and O-points (see upper right panel).
This fragmentation phase corresponds to a net increase of the resistive electric field contribution in the particle acceleration mechanism (see corresponding lower panel). 
The increase of the resistive contribution during the fragmentation phase may be explained by the formation of X-points, in which the resistive electric field is strong. 
During the fragmentation phase, plasmoids merge with each other and during this merging process the resistive contribution remains strong, reaching a peak at $\omega_p t = 4.0 \times 10^5$ (see lower left panels).
Indeed, when two plasmoids merge, a secondary current sheet forms at the interface between the two \citep[see, e.g.,][]{Oka2010, Sironi2014, Nalewajko2015}. 
Plasmoids merge until a large final magnetic island is formed.
When major mergers no longer occur, the resistive field contribution begins to smoothly decrease (see lower right panels).
Indeed, towards the end of the computational time, the resistive contribution seems to approach a saturation value, that is when the final giant plasmoid stabilizes \citep[called saturation phase in][]{Puzzoni2021}. 

Therefore, high-energy particles are accelerated by the resistive electric field when they cross an X-point \citep[and are shortly after injected into a plasmoid, see, e.g.,][]{Zenitani2001, Bessho2007, Lyubarsky2008, Sironi2014, Ball2019} and during islands merging. 
When particles are finally trapped inside the large final plasmoid, the resistive contribution decreases, leaving room for the more gradual action of the convective electric field.
These results are in agreement with those of \cite{Sironi2022}, who demonstrated that high-energy particles must have crossed non-ideal regions during the early stages (called \lq\lq injection\rq\rq\  by the author) of their acceleration process.
On the contrary, our results are in contrast with those of \cite{Guo2019}, who claim that the non-ideal field can be neglected in the particle acceleration mechanism, as the Fermi mechanism is the dominant one.
Similarly, our results are different from those of \cite{Kowal2011, Kowal2012} \citep[see also][]{deGouveia2015, delValle2016, Medina2021}, who argue that the resistive contribution is completely negligible in the particle acceleration mechanism.

\section{Summary}
\label{sec:summary}
%
%
In this work, we analyzed the role and importance of the resistive electric field in the process of particle acceleration in a reconnecting 2D Harris current sheet.
Our numerical simulations have been carried out with the PLUTO code for plasma astrophysics by simultaneously solving the non-ideal MHD equations with constant resistivity \citep[see][]{Mignone2007, Mignone2012} as well as the motion of test particles.
We choose a Lundquist number $\bar{S}=10^4$, and carry out computations at different grid resolutions ($N_x = 1536, 3072, 6144$).

Our results indicate clear convergence at intermediate energies ($10 \lesssim E_\mathrm{kin} \lesssim 10^3$), while at high energies ($E_\mathrm{kin} \gtrsim 10^3$) convergence achievement is not 
clear–cut.
However, even if the contribution of the resistive field slightly decreases with grid resolution (at high energies), a more detailed analysis reveals that its omission from the particle equations of motion leads to lower (within a factor of $10$) maximum energies and steeper cuts (with a power-law index $p \approx 1.8$ at the largest resolutions) in the particle energy spectra.
This behaviour remains essentially unaffected by grid resolution.

We found that the resistive contribution is strongest as the current sheet starts to fragment and plasmoids start to merge \citep[see][]{Puzzoni2021}.
During this phase, in fact, the resistive contribution sharply increases as a large number of X-points is created (where the resistive electric field is predominant).
The presence of X-points is indeed essential in producing abrupt acceleration of particles at this stage \citep[as found, e.g., by][]{Zenitani2001, Bessho2007, Lyubarsky2008, Sironi2014, Nalewajko2015, Ball2019}.
Moreover, particles energy is boosted also during plasmoids merging, due to the anti-reconnection electric field therein \citep[as demonstrated, e.g., by][]{Oka2010, Sironi2014, Nalewajko2015}. 

The resistive contribution gradually decreases as the system evolves towards the final saturation phase.
Then, as particles are trapped inside the largest magnetic island, they are accelerated by the $1^\mathrm{st}$-order Fermi mechanism in a contracting plasmoid \citep[as found, e.g., by][]{Drake2006, Drake2010, Kowal2011, Bessho2012, Petropoulou2018, Hakobyan2021}, which is dominated by the convective electric field.

In addition, we found particles gaining energy in the core of small plasmoids, where the resistive field is strong.  This is a feature that is missing in PIC simulations \citep[see, e.g.,][]{Petropoulou2018} and it may be an undesirable consequence of a constant resistivity approach. 
Future works will explore different resistivity models that could be more consistent in approaching collisionless plasmas.

Our results lead us to conclude that not only the resistive field is not-negligible \citep[in agreement with the works, for example, of,][]{Onofri2006,Zhou2016, Ball2019,Sironi2022}, but it plays a fundamental role in accelerating high-energy particles.
In particular, our results favourably agree with \cite{Sironi2022}, who argues that the non-ideal field is crucial in the early-stages of particle acceleration.
On the other hand, our outcomes disagree with those of \cite{Kowal2011, Kowal2012}, \cite{deGouveia2015}, \cite{delValle2016} and \cite{Medina2021}, who neglect the resistive electric field in the particle equations of motion as they do not consider it important in the acceleration process.
Similarly, our results also differ from those of \cite{Guo2019}, who argue that the Fermi mechanism is dominant and the non-ideal field can be neglected in the particle acceleration process during large-scale reconnection events, as it is unimportant for the formation of the power-law distribution.

Future simulations will address the issue of longer simulations and different choices of boundary conditions, as well as the extension to the relativistic regime.

\section*{Acknowledgements}
We acknowledge support by CINECA through the  Accordo Quadro INAF-CINECA for the availability of high-performance computing resources (project account INA21\_C8B67). 
We wish to thank Lorenzo Sironi and Bart Ripperda for their useful insights during the preparation of this manuscript.

\section*{Data Availability}
PLUTO is publicly available and the simulation data will be shared on reasonable request to the corresponding author.




\bibliographystyle{mnras}
\bibliography{paper} 



\appendix

\section{Plasma convergence study}
\label{app:plasma_convergence}

\begin{figure*}
    \centering
    \includegraphics[width=0.49\textwidth]{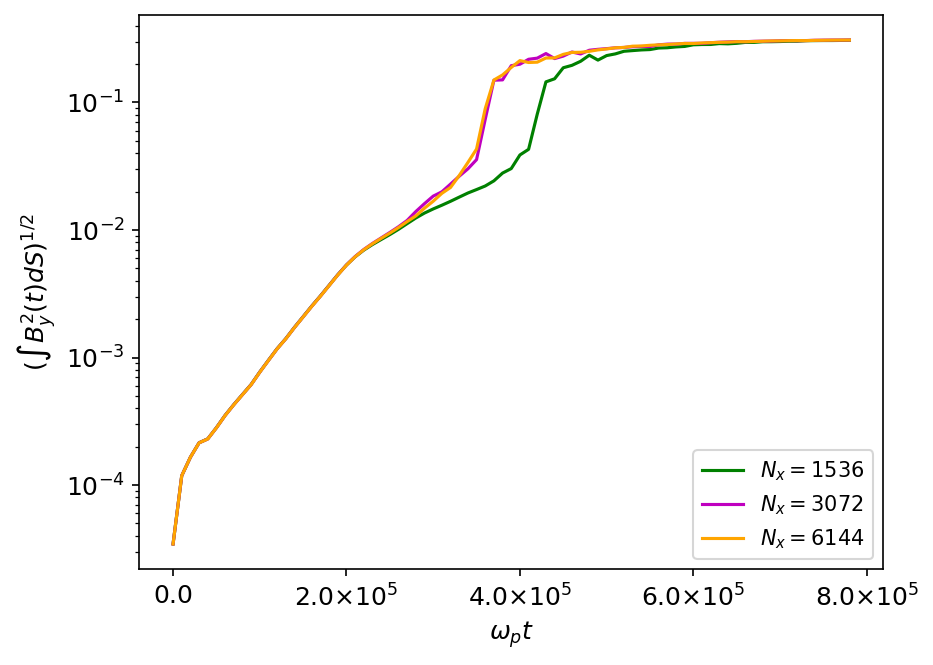}
    \includegraphics[width=0.485\textwidth]{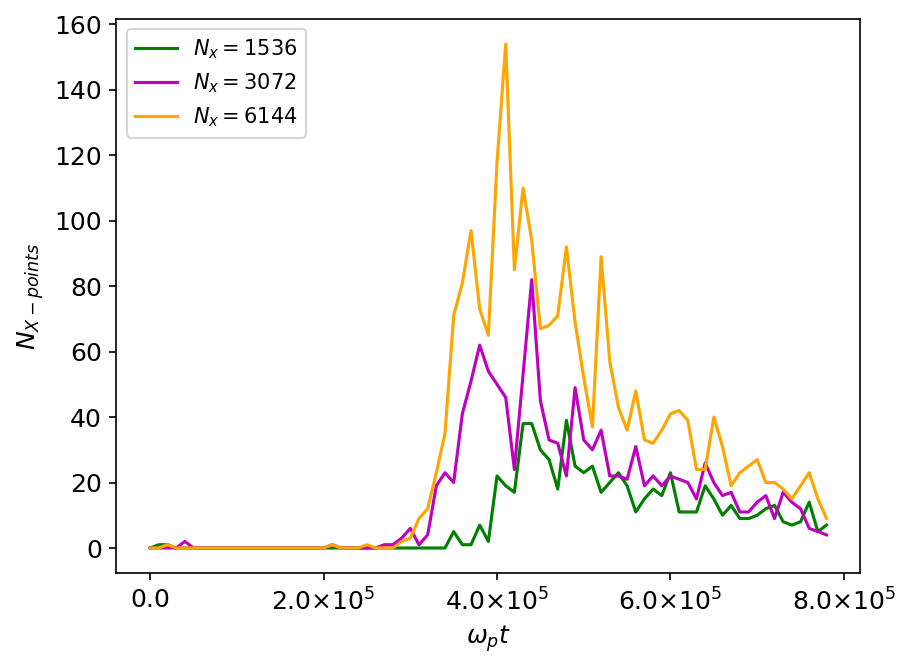}
    \caption{\textit{Left panel}: Spatially-averaged transverse component of magnetic field as a function of time at different grid resolutions. \textit{Right panel}: Number of X-points formed over time at the same grid resolutions.}
    \label{fig:plasma_convergence}
\end{figure*}

Here we focus on the plasma convergence study, following the methodology adopted in \cite{Puzzoni2021}.
Figure \ref{fig:plasma_convergence} shows the temporal evolution of the spatially-averaged transverse component of magnetic field at different grid resolutions (left panel) and the corresponding number of X-points formed (right panel).
The number of X-points is obtained through the algorithm based on locating the null points of $|\mathbf{B}|$ discussed in the Appendix of \cite{Puzzoni2021}.

Although the growth of the perturbation shown in the left panel seems to indicate convergence at $N_x = 3072$ even in the non-linear phase, we cannot conclude the same by looking at the right panel.
Indeed, the number of X-points increases with the grid resolution.
This leads us to conclude that, during the fragmentation phase, very thin current sheets are created, which are not completely resolved even at these high resolutions.
However, as emphasized in the paper, there are strong indications that our results are valid even if we have not achieved complete convergence.


\bsp	
\label{lastpage}
\end{document}